\documentclass[pra,twocolumn,showpacs,amsmath,amssymb]{revtex4}
\usepackage{graphicx}
\usepackage{amsfonts}
\usepackage{amsmath}
\usepackage{amssymb}
\usepackage{color}
\usepackage{bm}
\usepackage{bbm}
\usepackage{enumerate}

\graphicspath{{figures/}} 
\usepackage{amsfonts, amsmath, amsthm, amssymb} 
\usepackage{array}
\usepackage[colorlinks,bookmarks=false,citecolor=blue,linkcolor=red,urlcolor=blue]{hyperref}

\newcommand{\be}{\begin{equation}}
\newcommand{\bee}{\begin{equation*}}
\newcommand{\ee}{\end{equation}}
\newcommand{\eee}{\end{equation*}}
\newcommand{\bearre}{\begin{eqnarray*}}
\newcommand{\eearre}{\end{eqnarray*}}
\newcommand{\bearr}{\begin{eqnarray}}
\newcommand{\eearr}{\end{eqnarray}}

\usepackage{tikz}
\newcommand{\osum}{ 
  \mathop{
    \mathchoice
      {\buildosum{\displaystyle}{0.1}}
      {\buildosum{\textstyle}{0.075}}
      {\buildosum{\scriptstyle}{0.075}}
      {\buildosum{\scriptscriptstyle}{0.075}}
  }\displaylimits 
}

\newcommand\buildosum[2]{%
  \begin{tikzpicture}[baseline=(char.base), inner sep=0, outer sep=0]
    \draw (-0.3ex,0) circle (#2);
    \node (char) at (0,0) {$#1\sum$};
  \end{tikzpicture}%
}

\begin{document}

\title{Auxiliary-Field Monte Carlo for lattice bosons: tackling strong interactions and frustration}

\author{Daniele Malpetti$^1$
 and Tommaso Roscilde$^{1,2}$
 }

\affiliation{$^1$ Laboratoire de Physique, CNRS UMR 5672, Ecole Normale Sup\'erieure de Lyon, Universit\'e de Lyon, 46 All\'ee d'Italie, 
Lyon, F-69364, France}
\affiliation{$^2$ Institut Universitaire de France, 103 boulevard Saint-Michel, 75005 Paris, France}

\date{\today}

\begin{abstract}
We introduce a new numerical technique -- bosonic auxiliary-field Monte Carlo (bAFMC) -- which allows to calculate the thermal properties of large lattice-boson systems within a systematically improvable semiclassical approach, and which is virtually applicable to any bosonic model. Our method amounts to a decomposition of the lattice into clusters, and to an Ansatz for the density matrix of the system in the form of a cluster-separable state -- with non-entangled, yet classically correlated clusters. This approximation eliminates any sign problem, and can be systematically improved upon by using clusters of growing size. Extrapolation in the cluster size allows to reproduce numerically exact results for the superfluid transition of hardcore bosons on the square lattice, and to provide a solid quantitative prediction for the superfluid and chiral transition of hardcore bosons on the frustrated triangular lattice.     
\end{abstract}

\maketitle


\textit{Introduction}. Models of strongly correlated bosons on a lattice (or lattice-boson field theories) play a central role in the description of quantum many-body systems, encompassing the whole of quantum magnetism (due exact spin-boson mappings) \cite{Mendels-book,Diep2013} and including superconducting networks \cite{FazioZ2001} and ultracold bosons in optical lattices \cite{Blochetal2008,Krutitsky2016} to cite some relevant examples. Large-scale numerical approaches, particularly those based on quantum Monte Carlo (QMC) \cite{Sandvik2010,Pollet2012}, have been instrumental in the understanding of the equilibrium properties of quantum magnets and strongly correlated bosons (see Refs.~\cite{Zvyaginetal2007,Trotzkyetal2010} for some recent examples). Nonetheless, the presence of frustrated couplings in the magnetic Hamiltonians, or, more generally, of gauge fields in the lattice-boson Hamiltonians, leads inevitably to a well-known \emph{sign problem} for the QMC approach, which essentially prevents simulations from making any quantitative prediction in the relevant parameter regimes. Overcoming this limitation is an urgent problem, when considering the significant progresses in the experimental study of bosonic frustration with quantum magnets \cite{Mendels-book,Diep2013} or ultracold atoms in artificial gauge fields \cite{Dalibardetal2011,Goldmanetal2014}.

 In the face of the significant hurdles to simulate bosonic frustration, a valuable guiding principle to attack lattice bosonic field theories is to capture qualitative as well as quantitative traits of their physics using states which are \emph{weakly entangled} in real space. This principle is at the basis of two most common approaches to interacting bosons: 1) Gutzwiller mean-field (MF) theory \cite{Krutitsky2016,McIntoshetal2012,Luehmann2013}, used to predict phase diagrams of strongly correlated bosons, despite the fact that it eliminates any form of entanglement (as well as of correlation \emph{in toto}) between spatial building blocks (single sites or clusters thereof); 2) and $c$-field (CF) theory  \cite{c-field-book,Blakieetal2008}, which accounts at most for weak quantum effects, describing uniquely regimes which have a classical analog, but nonetheless incorporates fluctuations when supplemented with stochastic treatments such as Monte Carlo. Recently we have shown \cite{MalpettiR2016} that quantum many-body systems at finite temperature exhibit a strong spatial separation between quantum coherent fluctuations -- whose wavelengths are upper-bounded by a quantum coherence length $\xi_Q(T)$ which is finite as long as $T>0$ -- and thermal fluctuations, whose wavelengths can be arbitrarily large upon approaching a critical point. In particular, degrees of freedom separated by a distance larger than $\xi_Q$ are nearly separable: hence the system admits a description in terms of states which possess short-range entanglement only, but which can exhibit classical correlations of arbitrary range. Clearly one would need the complementary strengths of MF theory and CF theory to acquire a satisfactory description.  

 This letter introduces a new, semi-classical numerical method -- bosonic auxiliary-field Monte Carlo (bAFMC)-- which is precisely designed to exploit the separation of scales between quantum and classical fluctuations at finite temperature. bAFMC breaks a lattice boson or spin system into clusters which are treated exactly, and which are further coupled via a fluctuating classical auxiliary field (AF) mediating classical correlations. Quantum fluctuations are faithfully described up to the length scale of a cluster, while a Monte Carlo treatment of the AF allows to account for for thermal fluctuations to all length scales. The cluster decomposition introduces therefore an artificial cutoff in the wavelengths of quantum fluctuations, that can be removed via an extrapolation of the results to infinite cluster size. We validate our approach, showing that it can reproduce quantitatively the thermodynamics of a strongly quantum lattice-boson problem, namely the Berezhinskii-Kosterlitz-Thouless (BKT) transition of hardcore bosons on a square lattice; and we further apply it to reconstruct the phase diagram of hardcore bosons on the frustrated (or $\pi$-flux) triangular lattice.

 {\emph{Model Hamiltonian and path-integral treatment.} For the sake of concreteness, we shall focus on the case of the Bose-Hubbard model with arbitrary hopping terms
 \begin{equation}
 \hat{\cal H} = \sum_{i< j} \hat{h}_{ij} + \sum_i \hat{g}_i
 \label{e.Ham}
 \end{equation}
 where 
  $\hat{h}_{ij} = - J_{ij} \hat{b}_i^{\dagger} \hat{b}_j + {\rm h.c.}$ and $\hat{g}_i =  \frac{U}{2} (\hat{b}_i^+)^2 \hat{b}_i^2 - \mu \hat{b}_i^{\dagger} \hat{b}_i~$.
Here $\hat{b}_i, \hat{b}_i^{\dagger}$ are bosonic operators, and the indices $i$ and $j$ run on the sites of a $d$-dimensional lattice. In the most general case the matrix $J_{ij}$ is hermitian, and its complex matrix elements describe the presence of a gauge field. While motivated by the field of cold atoms \cite{Blochetal2008,Krutitsky2016,Goldmanetal2014} this model is also of immediate relevance to (frustrated) quantum magnetism when taking the limit $U\to \infty$, which produces a quantum $S=1/2$ XY model. \footnote{The following discussion can be readily generalized the presence of arbitrary diagonal interactions, as well as to the XXZ model of magnetism \cite{MalpettiR2016b}.}.

\begin{figure}
 \includegraphics[width = \linewidth]{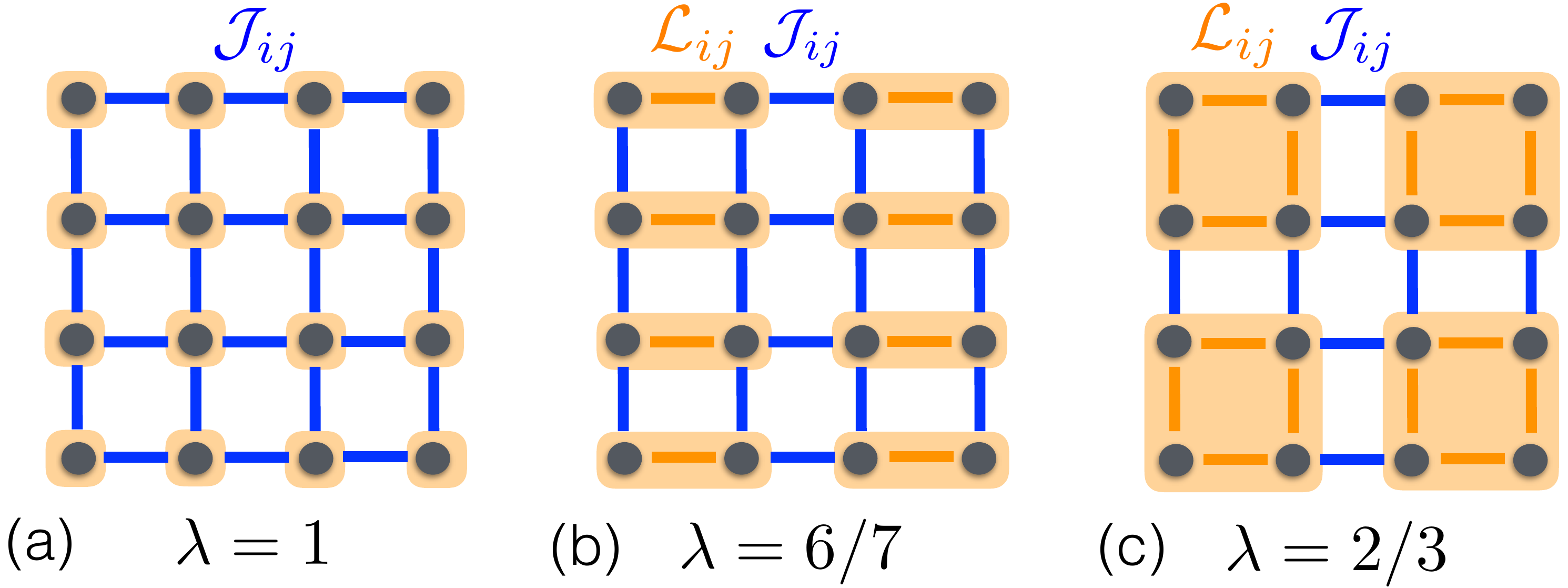}
	\caption{Cluster decompositions of the square lattice lattice with nearest-neighbor hoppings. Orange-shaded areas identify the clusters, with intracluster bonds ${\cal L}_{ij}$ marked in orange; the inter-cluster bonds ${\cal J}_{ij}$ are instead marked in blue. The $\lambda$ parameter is the surface-to-bulk ratio of the cluster (see text).}
	\label{f.cluster}
      \end{figure}

 Our approach starts by decomposing the lattice into clusters (see Fig.~\ref{f.cluster}), implying a decomposition of the hopping matrix as $J_{ij} = {\cal J}_{ij} + {\cal L}_{ij}$, where ${\cal J}_{ij}$ is the matrix of the \emph{inter}-cluster couplings, while ${\cal L}_{ij}$ contains only the \emph{intra}-cluster couplings. The path-integral treatment of the Bose-Hubbard model \cite{Fisheretal1989} allows to decouple the various clusters via a Hubbard-Stratonovich (HS) transformation introducing an imaginary-time-dependent, complex auxiliary field $\Phi_i(\tau)$, which is defined on the sets ${\cal C}_c$ of  "boundary" sites $i$ of the clusters (labeled by the $c$ index), satisfying the condition that ${\cal J}_{ij}\neq 0$ for some $j$. This leads then to the following form for the partition function (see Supplementary Material (SM) for an explicit derivation \cite{suppmat}):
 \begin{equation}
 {\cal Z} = \int {\cal D}[\Phi_{i}(\tau)] ~\exp(-S[\Phi_i(\tau)]) ~\prod_c {\cal Z}_c[\{\Phi_{i\in {\cal C}_c} (\tau)\}]
 \label{e.Z}
 \end{equation}
where 
\begin{equation}
S[\Phi_i(\tau)] = \int d\tau \sum_{ij} \Phi_i^*(\tau) (\tilde{\cal J}^{-1})_{ij} \Phi_j(\tau)
\end{equation}
is the action involving exclusively the auxiliary fields, while 
\begin{equation}
{\cal Z}_c[\Phi_{i \in {\cal C}_c}(\tau)] = {\rm Tr}\left [T_{\tau} ~e^{-\int d\tau \hat{\cal H}_c(\{\Phi_i(\tau),\Phi^*_i(\tau)\})} \right]
\end{equation}
is the effective partition function of a single cluster: here $T_{\tau}$ is the imaginary-time ordering operator, and  $\hat{\cal H}_c (\{\Phi_i(\tau),\Phi^*_i(\tau)\}) = \sum_{i,j \in c} \hat h_{ij} -  \sum_{i\in {\cal C}_c} (\Phi_i(\tau) \hat b_i^{\dagger} + \Phi^*_i(\tau) \hat  b_i) + \sum_{i \in c} (\hat{g}_i + K \hat{n}_i )$ is the effective single-cluster Hamiltonian, including the intra-cluster hopping, the coupling to the auxiliary field and the local diagonal terms. Moreover we have introduced the shifted matrix $\tilde{\cal J}_{ij} = {\cal J}_{ij} + K \delta_{ij}$, with $K = (1+\epsilon) |\Lambda_{\rm min}|$ and $\Lambda_{\rm min}$ the minimal (negative) eigenvalue of ${\cal J}$; an $\epsilon >0$  assures positive-definiteness of $\tilde{\cal J}$, as required by the HS transformation. The shift $K$ is then compensated by a complementary shift in the chemical potential appearing in $\hat{\cal H}_c$.\footnote{The chemical potential shift only affects the sites $i$ within a cluster which are coupled via ${\cal J}_{ij}$ to a neighboring cluster; in the case of short-range couplings, the chemical potential shift occurs therefore only on the boundary of each cluster.}

 \emph{Quantum mean-field approximation and auxiliary-field Monte Carlo.} The expression Eq.~\eqref{e.Z}  for the partition function (widely used as a basis of the field-theoretical treatment \cite{Fisheretal1989,Herbutbook}) is exact, but impractical for a Monte Carlo sampling, since the single-cluster partition functions ${\cal Z}_c$ are generally complex objects, leading to a sign problem \cite{UlmkeS2000}. To cast the AF formulation of the partition function into a practical tool for numerics, an approximation is in order. A most natural one - turning ${\cal Z}_c$ into a positive real number - is to treat the AF as a \emph{classical} complex field, namely $\Phi_i(\tau) = \Psi_i$ independent of $\tau$. Such an approximation amounts to decoupling clusters in their imaginary-time fluctuations: as discussed in Refs.~\cite{MalpettiR2016,MalpettiR2016b}, this is equivalent to decoupling their quantum fluctuations via a so-called cluster \emph{quantum} mean-field (cQMF) approximation (namely a mean-field approximation restricted to quantum fluctuations only). This corresponds to casting the density matrix $\hat \rho$ of the system (such that ${\cal Z} = {\rm Tr} \hat \rho$) into the form:
 \begin{equation}
 \hat\rho \approx \hat\rho_{\rm cQMF} = \int  {\cal D}[\Psi] ~P[\Psi]~ \otimes_c \hat\rho_c(\{\Psi_{i\in {\cal C}_c}, \Psi^*_{i\in {\cal C}_c}\}) 
 \label{e.rho}  
 \end{equation}
 where ${\cal D}[\Psi] = \prod_{i\in \cal C} \frac{d{\Psi}_i d{\Psi}_i^*}{2\pi i}$ is the AF metric, $P[\Psi] = ({\rm det} X)^{-1} \exp[-\beta \sum_{ij} \Psi_i^* X_{ij} \Psi_j] $ and 
 $\hat\rho_c = \exp[-\beta {\cal H}_c(\{\Psi_i,\Psi^*_i\})]$. We have introduced the symbol $X = \tilde{\cal J}^{-1}$. Eq.~\eqref{e.rho} is easily recognizable as a \emph{separable} form for the density matrix \cite{Werner1989}, in which entanglement between clusters is absent; Eq.~\eqref{e.rho} actually expresses a strong form of separability, called Hamiltonian separability \cite{MalpettiR2016}, which implies absence of entanglement \emph{and} quantum correlations, while still describing classical correlations (according to the definition of Ref.~\cite{Werner1989}). 
 
 The partition function descending from the cQMF approximation,   ${\cal Z} \approx {\rm Tr} (\hat\rho_{\rm cQMF})$, describes then an effective classical field theory for the AF, governed by the action 
 \begin{equation}
 S_{\rm eff}[\Psi] =  \beta \sum_{ij} \Psi_i^* X_{ij} \Psi_j  - \sum_c \log {\cal Z}_c[\{\Psi_{i\in {\cal C}_c}\}]~.
 \label{e.Seff}
 \end{equation}
 This effective classical field theory results from integrating quantum fluctuations with wavelengths upper bounded by the linear size of the clusters, $l_c$. In the spirit of a real-space renormalization group transformations, this latter scale can be seen as a moving cutoff, setting the boundary between the fully quantum and the effective classical description of the system. By sending $l_c$ to infinity we recover the exact description of the system: as shown in Ref.~\cite{MalpettiR2016b}, a quantitative extrapolation of the cQMF results towards the exact description can be achieved as a power law in the bulk-to-boundary ratio $\lambda = N_{\rm ext}/(N_{\rm int} + N_{\rm ext})$, where $N_{\rm int}$ is the number of internal bonds to each cluster, while $N_{\rm ext}$ is the number of bonds connecting the cluster to the outside. The introduction of a cutoff scale for quantum fluctuations and entanglement is fundamentally justified at finite temperature by the finiteness of the quantum coherence length $\xi_Q$ \cite{MalpettiR2016}, beyond which two degrees of freedom can be considered as essentially (Hamiltonian) separable. The quality of the cQMF approximation is therefore controlled by the ratio between the two length scales $l_c$ and $\xi_Q$ \cite{MalpettiR2016b}. Finally it can be shown \cite{suppmat} that a \emph{saddle-point} approximation to the effective action, Eq.~\eqref{e.Seff}, reproduces cluster MF (cMF) theory (albeit with modified couplings and chemical potential). Hence the cQMF approximation is a clear improvement over cMF theory via the inclusion of inter-cluster classical correlations.
 
The bAFMC approach amounts to solve numerically the effective classical field theory, described by the action $S_{\rm eff}[\Psi]$, via Monte-Carlo sampling (see SM \cite{suppmat} for a detailed discussion). At zero temperature the saddle-point approximation to the classical auxiliary field becomes exact, so that in this limit the bAFMC approach reduces to a modified cMF theory \cite{suppmat}. Yet the finite-temperature behavior is captured by bAFMC beyond any mean-field description. Indeed the effective action $S_{\rm eff}$ possesses all the symmetries of the original Hamiltonian \footnote{The $U(1)$ invariance of the quadratic part is manifest, as well as that of the part containing the cluster partition functions ${\cal Z}_c$, given that the phase of the bosonic field which couples to that of the AF is traced over.}, and it preserves the short-ranged nature of the original couplings \cite{suppmat}. Therefore, unlike in any mean-field approach, a Monte Carlo sampling of the fluctuations governed by $S_{\rm eff}[\Psi]$ shall reproduce the correct nature of phase transitions or extended critical phases that one may expect in the original system.  

\begin{figure}
 \includegraphics[width = \linewidth]{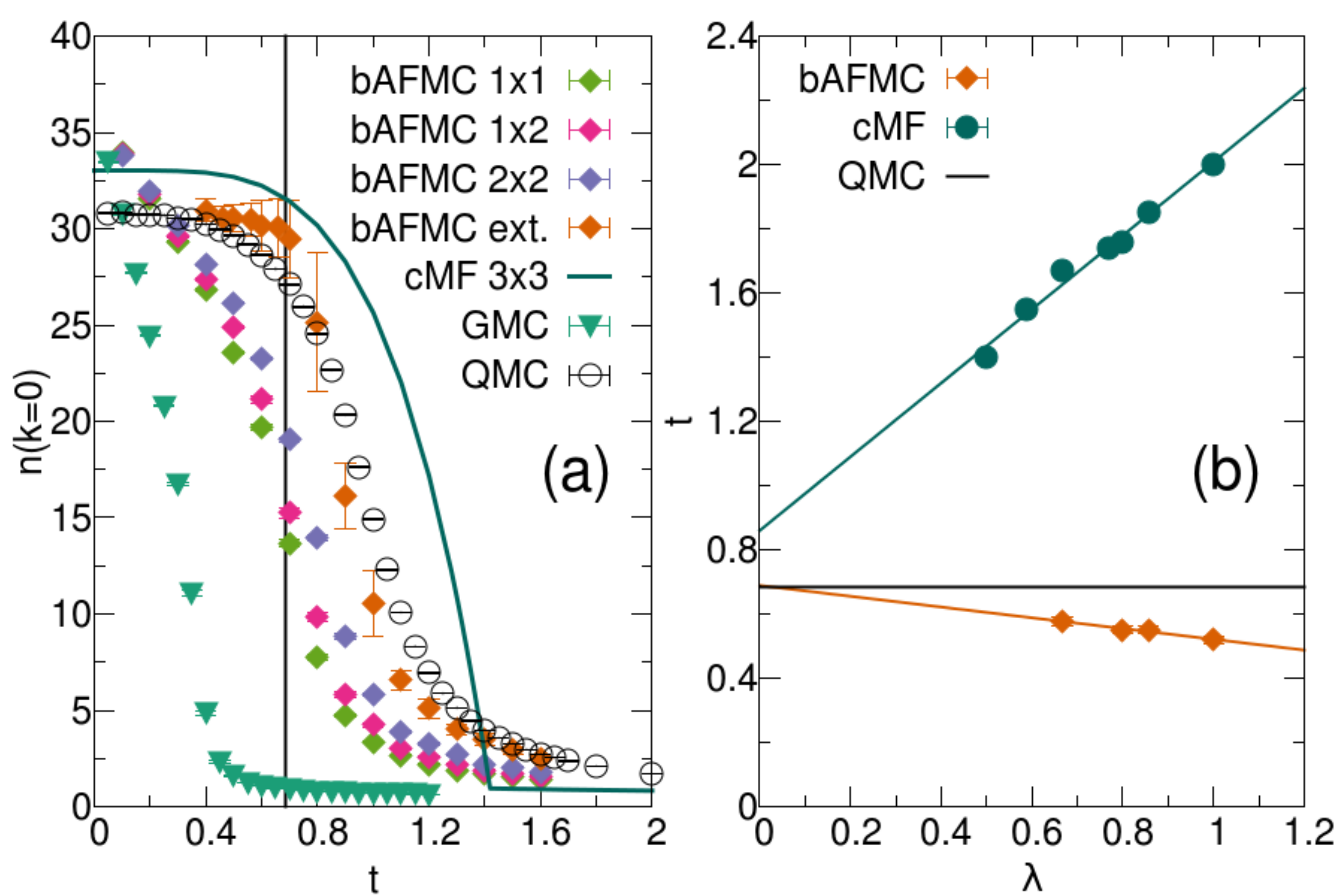}
	\caption{(a)  Comparison of exact QMC, bAFMC (for various cluster sizes, plus extrapolation), GMC and cMF results for $n(k=0)$ of hardcore bosons on a 12$\times$ 12 lattice; (b) cluster scaling of the BKT transition temperature from bAFMC, compared to the critical temperature from cMF; solid lines are linear fits, whose extrapolated $\lambda \to 0$ value is to be compared with the QMC value $t = 0.6854$ \cite{KawashimaH1998} (solid horizontal line).}
	\label{f.SL}
      \end{figure} 

\emph{Hardcore bosons on the square lattice.} As a first validation stage, we test the bAFMC approach in the case of hardcore bosons on the square lattice at half filling, corresponding to the quantum $S=1/2$ XY model on the same lattice. The Hamiltonian is readily obtained as a limiting case of Eq.~\eqref{e.Ham} with  $U\to \infty$, $\mu = 0$ and $J_{ij} = J$ for $i,j$ nearest neighbors on the square lattice, and zero otherwise.  We introduce the reduced temperature $t =k_B T/J$. This Hamiltonian features a BKT transition at 
 $t_{\rm BKT} \approx 0.6854$ (estimated via QMC) \cite{KawashimaH1998}, and an extended critical phase for $t<t_{\rm BKT}$, which are both inaccessible to mean-field treatments. Moreover the hardcore limit, while numerically favorable due to the restricted dimensions of the local Hilbert space, is the farthest possible from the classical limit of bosonic theories, and therefore possibly the hardest to describe quantitatively within a semi-classical setting. 

Fig.~\ref{f.SL}(a) shows the temperature dependence of the $k=0$ peak in the momentum distribution, $n(k=0) = \sum_{ij} \langle \hat{b}_i^{\dagger} \hat{b}_j \rangle / L^2$ for a lattice of size $L=12$, as obtained via different methods: 1) numerically exact QMC \cite{SyljuasenS2002}; 2) the cMF approach based on a $3\times 3$ cluster \cite{Luehmann2013}; 3) the semi-classical approach of Ref.~\cite{HickeyP2014} (here dubbed Gutzwiller Monte Carlo - GMC), which amounts to a Monte Carlo sampling of different Gutzwiller mean-field wavefunctions $|\Psi\rangle = \otimes_i |\psi_i\rangle$ weighted by the Boltzmann weight $e^{-\beta \langle \Psi | \hat{\cal H} |\Psi \rangle}$; and 4) the bAFMC
approach based on clusters of growing size from $1\times1$ up to $2\times2$.  The latter two approaches have the common aspect of reducing to cMF theory at zero temperature (albeit a modified one in the case of bAFMC \cite{suppmat}). The cMF predicts an unphysical true condensation transition for a 2$d$ system, whose temperature grossly overestimates the BKT temperature, and even an extrapolation in the size of the cluster turns out to be problematic (see below); on the opposite front, the Gutzwiller MC approach, while capturing correctly the BKT physics \cite{HickeyP2014}, significantly underestimates the transition, without offering any viable (\emph{e.g.} cluster-based) strategy for further improvement. The bAFMC results, on the other hand, are the closest ones to QMC among the three approximation schemes considered here: even though the considered cluster decompositions give results which remain relatively far from the exact ones, a clear trend towards the exact values is observed upon increasing the cluster size.  In particular, a systematic linear extrapolation in the $\lambda$ coefficient can be made which reproduces quite closely the exact results -- we would like to stress that the residual discrepancy is a limitation of the very basic linear extrapolation scheme (imposed by the limited number of cluster sizes we considered), and it can still be systematically improved upon. Most importantly, irrespective of the cluster size all effective classical theories produced by the bAFMC approach possess a genuine BKT transition, whose critical temperature can be estimated from the expected critical scaling $n(k=0) \sim L^{7/4}$  (here for system sizes $L=12, 24$ and $36$ \cite{suppmat}). The BKT temperature so extracted are then plotted as a function of the $\lambda$ parameter in  Fig.~\ref{f.SL}(b): a simple linear extrapolation towards $\lambda = 0$ produces the estimate $t_{\rm BKT}(\lambda = 0) = 0.69(2)$, in very good agreement with the QMC estimate. A similar extrapolation of the critical temperature for the cMF condensation transition does \emph{not} converge towards the QMC estimate, suggesting that, even within a cluster approach, the MF transition cannot be reliably used as an estimate of the quasi-condensation transition of 2$d$ hardcore bosons. 

\begin{figure}
 \includegraphics[width = \linewidth]{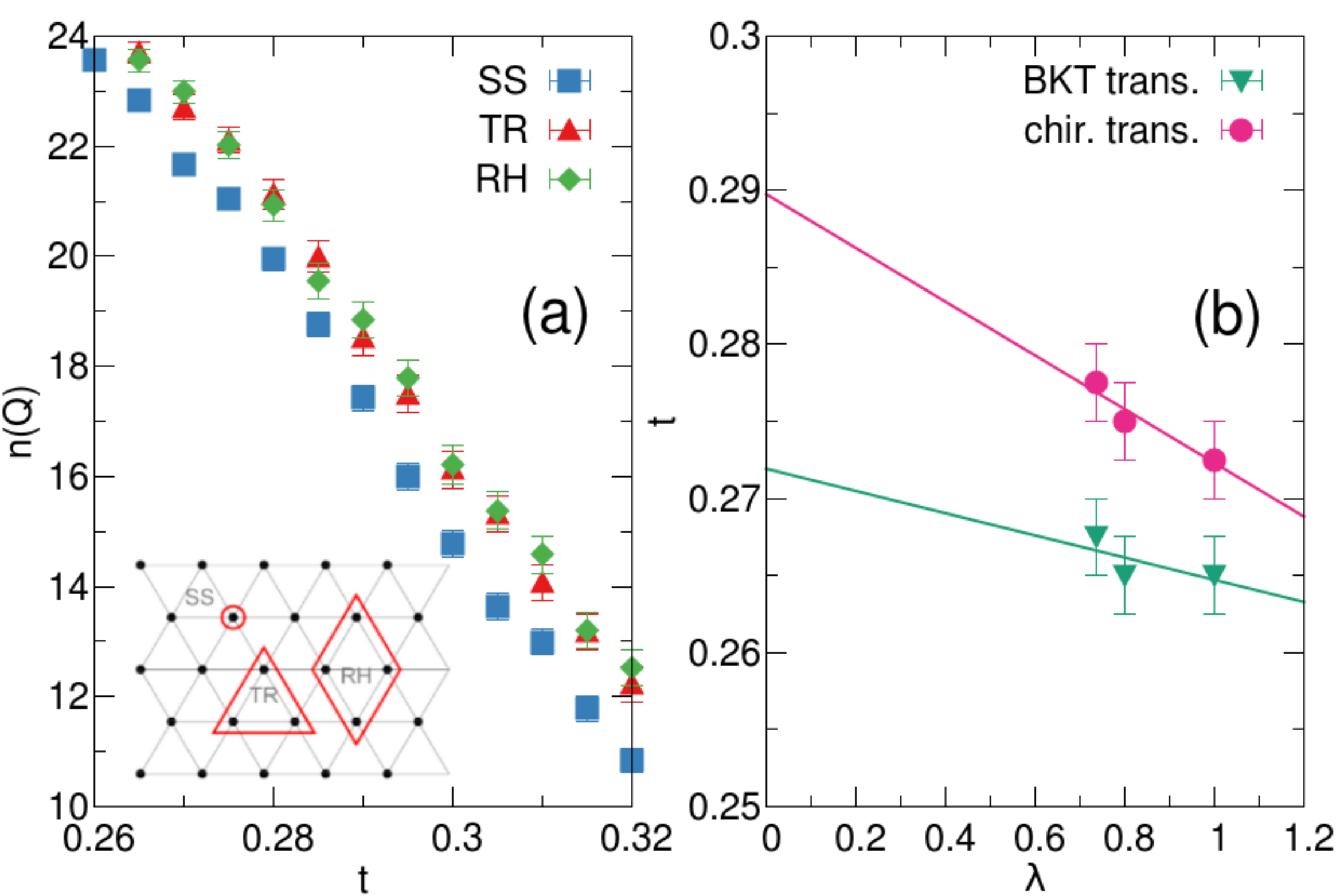}
	\caption{(a) Momentum distribution peak $n({\bm Q})$ of hardcore bosons on the 12$\times$12 frustrated triangular lattice from bAFMC on different cluster sizes (pictured in the inset); (b) Cluster scaling of the BKT and chiral transition temperatures; solid lines are linear fits.}
	\label{f.TL}
      \end{figure} 

\emph{Hardcore bosons on the triangular lattice.}  Having validated the bAFMC for hardcore bosons on the square lattice, we move on to apply it to an open problem of strongly correlated bosons in the presence of a frustrating gauge field, namely the case of hardcore bosons on a triangular lattice (TL) with a $\pi$-flux (or Eq.~\eqref{e.Ham} $J_{ij} = -J$ for nearest-neighbor sites, and other parameters as for the square lattice), corresponding to the antiferromagnetic $S=1/2$ XY model on the same lattice. The ground state of the model displays three-sublattice long-range order, which entails the ordering of both the spin variables (bosonic phases) as well as of the spin chirality (bosonic plaquette current) $\kappa_{\triangle}  = \osum (i\hat b_i^{\dagger} \hat b_j + {\rm h.c.})$, where the sum runs over oriented $ij$ pairs on the elementary triangular plaquette. In the classical spin ($S\to\infty$) limit a broad consensus exists \cite{ObuchiK2012} about the existence of two thermal phase transitions: a lower-temperature BKT transition at $T_{\rm BKT}$ with quasicondensation at finite momentum on the corners of the Brillouin zone ($\pm{\bm Q} = \pm(4\pi/3,0)$ and equivalent wavevectors), and a higher temperature \emph{chiral} transition at $T_c$ associated with the appearance of a vortex lattice: the latter is characterized by a divergence of the chirality structure factor $S_{\kappa} = L^{-2} \sum_{\triangle,\triangle'} \langle  \kappa_{\triangle} \kappa_{\triangle'} \rangle$. Chiral ordering on the triangular lattice has been recently observed by cold-gas experiments in the bosonic classical limit (large occupancy with weak interactions) \cite{Strucketal2011, Strucketal2013}.
 
 Our bAFMC investigation of the frustrated TL focused on three different cluster shapes: single site, triangular and rhombic (see Fig.~\ref{f.TL}(a)). The temperature dependence of the momentum distribution peak $n({\bm Q})$ (shown in Fig.~\ref{f.TL}(a)) as well as that of the chirality structure factor $S_{\kappa}$ (see \cite{suppmat}) are observed to depend rather weakly on the cluster shape around the BKT and chiral transitions : this is a clear signature that the range of quantum correlations in the thermal critical regime is strongly suppressed by frustration with respect to the case of the unfrustrated square lattice. Correspondingly the estimates of the critical temperatures $T_{\rm BKT}$ and $T_c$ extracted from finite-size scaling show a rather weak dependence on the $\lambda$ parameter (see Fig.~\ref{f.TL}(b)), which gives further confidence in their extrapolation to $\lambda\to 0$. The separation between $T_{\rm BKT}$ and $T_c$ increases when $\lambda$ decreases, and their extrapolated values ($T_{\rm BKT}(\lambda = 0) = 0.272(7)$,  $T_c(\lambda = 0) = 0.290(4)$) exhibit a sizable relative separation of 6\%, to be compared with the separation of 2\% in the classical spin limit  \cite{ObuchiK2012}. This shows that quantum effects can strongly increase the delicate spin-chirality decoupling observed in classical frustrated magnets, possibly to a level at which it becomes observable with state-of-the-art experiments on cold-atom quantum simulators. \footnote{Our critical temperature estimates are in sharp disagreement with the previous estimates present in the literature ($T_{\rm BKT} = 0.375(1)$ and $T_c=0.386(4)$ from Ref.~\cite{Capriottietal1999}): it should be noticed nonetheless that the latter are based on a semiclassical approach to the spin-$S$ quantum XY model which is not well controlled in the $S=1/2$ limit, and which may systematically underestimate quantum effects and hence overestimate the critical temperatures.}  
 
 \emph{Conclusions.} We have introduced a new numerical approach (the bosonic auxiliary-field Monte Carlo) based on a semiclassical approximation to the partition function which eliminates any sign problem at the expense of the truncation of long-range quantum correlations beyond a given cutoff, set by a cluster decomposition of the lattice. This approximation is well controlled due the generically short-ranged nature of quantum correlations at finite temperature, and most importantly it can be systematically improved by moving the cutoff to larger length scales. Our approach positions itself among the methods which are limited by entanglement and quantum correlations -- such as tensor-network Ans\"atze, including the density-matrix renormalization group \cite{Orus2014}: while the success of the latter is mostly based on the weakness of entanglement in the ground state of physical Hamiltonians of interest, the success of our method exploits for the first time the weak entanglement and quantum correlations present in \emph{thermal} states. Access to larger clusters than the ones used here could be easily granted by the use of Lanczos reconstruction of the low-lying spectrum \cite{Prelovsek2013} (when treating sufficiently low temperatures) or by the recently proposed  reconstruction of an effective auxiliary-field Hamiltonian from a limited sample of configurations \cite{Liuetal2016}. The wide applicability of our approach to bosonic systems makes it a very suitable candidate to investigate prominent models of frustration, which are of central interest to quantum magnetism and quantum simulation with ultracold atoms. 
 
 \emph{Acknowledgements.} We thank I. Fr\'erot, A. Ran\c con and P. Naldesi for useful discussions. This work is supported by the Programme Avenir Lyon Saint-Etienne (ANR-11-IDEX-0007) of Universit\'e de Lyon within the program ``Investissements d'Avenir" (ANR), and by ANR JCJC-2013 (``ArtiQ" project).  All simulations were performed on the PSMN cluster of the ENS of Lyon.

\newpage
\newpage

\begin{widetext}
 \begin{center}
 {\bf SUPPLEMENTARY MATERIAL for: \emph{Auxiliary-Field Monte Carlo for lattice bosons: tackling strong interactions and frustration}}
 \end{center}
 
 \section{Auxiliary-field formulation of the density matrix and partition function}
 
 The partition function of strongly interacting lattice bosons can be expressed as an integral over the auxiliary fields $\{\Phi_i(\tau), \Phi^*_i(\tau)\}$ making use of the coherent-state path-integral approach, as pioneered in Ref.~\cite{Fisheretal1989}. Nonetheless coherent-state path integrals for bosonic as well as spin systems have recently come under scrutiny, because they are found lead to erroneous results when calculated in the strict continuous-time limit, due to the overcomplete nature of the coherent-state basis \cite{WilsonG2011,Yanay2015}. This is not at all an issue for our formulation, given that, after Hubbard-Stratonovich decoupling, the coherent-state path integral is recast in an operator form to give the single-cluster partition functions ${\cal Z}_c$, eliminating any ambiguity. To further corroborate this statement, we show here that one can derive the auxiliary-field formulation of the partition function working uniquely with quantum operators, namely without making use of coherent states.  
 
 A central formula which shall be useful in the following involves Gaussian integrals of complex variables
 \begin{eqnarray}
 e^{-\Delta \tau [-\Gamma ~\hat b^{\dagger} \hat b  + \hat O(\hat b,\hat b^{\dagger})]} & = & 1 + \Delta \tau [\Gamma ~\hat b^{\dagger} \hat b  - \hat O(\hat b,\hat b^{\dagger})] +  ~{\cal O}(\Delta \tau^2) \nonumber \\ 
 & = & \int  \frac{d\Phi d\Phi^*}{2\pi i \Gamma} e^{-\Delta\tau |\Phi|^2/\Gamma} ~ \left [1 + \Delta \tau (\Phi^* \hat b + \Phi \hat b^{\dagger} - \hat O) 
 + \frac{\Delta \tau^2}{2}  \left(\Phi^* \hat b + \Phi \hat b^{\dagger}\right)^2 + \frac{\Delta \tau^2}{2}  |\Phi|^2  +  ~{\cal O}(\Delta \tau^2) \right ] \nonumber \\
 & = & \int  \frac{d\Phi d\Phi^*}{2\pi i \Gamma} e^{-\Delta\tau |\Phi|^2/\Gamma} ~ \left [1 + \Delta \tau (\Phi^* \hat b + \Phi \hat b^{\dagger} - \hat O) 
 + \frac{\Delta \tau^2}{2}  \left(\Phi^* \hat b + \Phi \hat b^{\dagger} + \hat{O} \right)^2 + \frac{\Delta \tau^2}{2} |\Phi|^2 + ~{\cal O}(\Delta \tau^2) \right ] \nonumber \\
  &=&   \int  \frac{d\Phi d\Phi^*}{2\pi i \Gamma} e^{-\Delta\tau \left (\frac{1}{\Gamma} - \frac{\Delta\tau}{2}\right ) |\Phi|^2} \left[ e^{ - \Delta \tau [-\Phi^* \hat b - \Phi \hat b^{\dagger} + \hat O(\hat b,\hat b^{\dagger})]}~ +~  {\cal O}(\Delta \tau^2) \right ]
 \label{e.formula}
 \end{eqnarray}
where $\hat O(\hat b,\hat b^{\dagger})$ is an arbitrary function of $\hat b$ and $\hat b^{\dagger}$; $\Delta \tau = \beta/M$ corresponds to a Trotter discretization of the imaginary time axis, with the assumption that the limit $M\to \infty$ shall be taken at the end of the calculation. The third and fourth terms on the second line are only apparently of order $\Delta \tau^2$, given that they contain a term $|\Phi|^2$ which is ${\cal O}(\Delta\tau^{-1})$ upon Gaussian integration.
 
We then write the Hamiltonian of the Bose-Hubbard model in the cluster-decomposed form 
\begin{eqnarray}
\hat{\cal H} & = & - \sum_{c<c'} \sum_{i\in {\cal C}_c} \sum_{j\in {\cal C}_{c'}}  \left ( \tilde{\cal J}_{ij} \hat{b}_i^{\dagger} \hat{b}_j  + {\rm h.c.} \right ) + \sum_c \left [ -\sum_{i,j \in c} \left ( J_{ij} \hat{b}_i^{\dagger} \hat{b}_j + {\rm h.c.} \right ) + \sum_{i \in c} \hat{g}_i(U,\mu_i) \right]  \nonumber \\
& = & \sum_\alpha (\Lambda_{\alpha} + K) ~\hat{b}_{\alpha}^{\dagger} \hat{b}_{\alpha} + \sum_c ~\hat{\cal K}_c
\end{eqnarray} 
where $\hat{g}_i$ is the single-site term containing the on-site interaction and the chemical potential term, which is shifted from to $\mu_i = \mu - K$ if site $i$ is coupled to sites outside its cluster by the matrix ${\cal J}_{ij}$, otherwise $\mu_i = \mu$. In the second line we have formally regrouped the intra-cluster terms into $\hat{\cal K}_c$, and transformed the ``boundary" field operators $\hat{b}_i, \hat{b}^{\dagger}_i$ (attached to $i$ sites which are coupled by the inter-cluster couplings) to the basis -- indexed by $\alpha$ -- which diagonalizes the matrix ${\cal J}_{ij}$, giving eigenvalues $\Lambda_{\alpha}$. Once again, the $K$ shift guarantees that $\Lambda_{\alpha} + K > 0$ for all $\alpha$. 
 
Making use of the above formula, Eq.~\eqref{e.formula}, we obtain for the density matrix 
\begin{equation}
\hat \rho = \lim_{M\to \infty} \left ( e^{-\Delta \tau \hat{\cal H}} \right )^M = \lim_{M\to \infty} \left [  \int  \left ( \prod_{\alpha,k} \frac{d\Phi_{\alpha,k} d\Phi^*_{\alpha,k}}{2\pi i (\Lambda_{\alpha}+K)} \right) e^{-\Delta\tau \sum_{\alpha,k} \left( \frac{1}{\Lambda_{\alpha}+K} -\frac{\Delta\tau}{2}\right) |\Phi_{\alpha,k}|^2}   \otimes_c \hat\rho_c  + {\cal O}(M \Delta \tau^2) \right ] ~.
\label{e.rho2}
\end{equation}
where
\begin{equation}
\hat \rho_c = \prod_{k=1}^M \exp \left \{ - \Delta \tau \left [-{\sum_{i\in c}}' \left( \Phi^*_{i,k} \hat b_i + \Phi_{i,k} \hat b_i^{\dagger} \right) + \hat{\cal K}_c \right ] \right \}~
~ \rightarrow ~ T_{\tau} ~\exp \left \{-\int d\tau \left[ -{\sum_{i\in c}}' \left( \Phi^*_i(\tau) \hat{b}_i + \Phi_i(\tau) \hat b^{\dagger} \right) + \hat{\cal K}_c \right ] \right \} 
\end{equation}
Here the $\sum'$ runs over the $i$ sites of cluster $c$ coupled to sites outside the cluster, and in the last step we have taken continuous-time limit $\Phi_{i,k} \to \Phi_i(\tau)$. The normal ordering of the infinitesimal imaginary-time evolution operator has essentially no effect, given that the $\hat{\cal K}_c$ is already normally ordered.  
Moreover 
\begin{equation}
\Delta\tau \sum_{\alpha,k} \left( \frac{1}{\Lambda_{\alpha}+K} -\frac{\Delta\tau}{2}\right) |\Phi_{\alpha,k}|^2 ~ \rightarrow ~
-\int d\tau \sum_{ij} \Phi_i^*(\tau) X_{ij} \Phi_j(\tau)   ~.
\end{equation}
Therefore the $M\to\infty$ limit of Eq.~\eqref{e.rho2} delivers the path-integral form over time-dependent auxiliary fields $\Phi_i(\tau)$ for the density matrix and the partition function reported in the main text.
 
 \section{Zero-temperature limit of bosonic auxiliary-field Monte Carlo vs. cluster mean-field theory}
 
 At zero temperature the bAFMC approach reconstructs the auxiliary field configuration $\{\Psi_i\}$ minimizing the effective energy 
 \begin{eqnarray}
 && E[\Psi_{i},\Psi^*_{i}] = \lim_{\beta\to\infty} S_{\rm eff}/\beta \nonumber \\
 &=&  \sum_{ij} \Psi_i^* X_{ij} \Psi_j  + \sum_c \langle \hat{\cal H}_c[\Psi_{i\in c},\Psi^*_{i\in c}] \rangle_0
 \end{eqnarray} 
 where $\langle ... \rangle_0$ defines the expectation value on the ground state of ${\cal H}_c$, which is in turn a function of the auxiliary field. Minimizing with respect to the auxiliary field leads to the condition 
 \begin{equation}
 \frac{\delta}{\delta \Psi_i^*} E[\Psi_{i},\Psi^*_{i}]  = -\langle  \hat{b}_i \rangle_0 + \sum_j X_{ij} \Psi_j = 0
 \end{equation}
 which translates into a self-consistent equation for the ground-state auxiliary fields: 
 \begin{equation}
 \Psi_i = \sum_j {\cal J}_{ij} \langle \hat{b}_j \rangle_0 +  K \langle \hat{b}_i \rangle_0
 \end{equation}
 Upon setting $K=0$  (both explicitly in the above equation, as well as inside the effective cluster Hamiltonian $\hat{\cal H}_c$ whose chemical potential is shifted by $-K$) we would recover the self-consistent equation of cluster mean-field (cMF) theory. On the other hand the requirement of positive definiteness of the matrix $\tilde{\cal J}_{ij}$ imposes that $K = (1+\epsilon)|\Lambda_{\rm min}| > 0$.  This gives to the zero-temperature limit of bAFMC the structure of the cMF solution of a \emph{modified} model, with a shifted chemical potential for the ``boundary sites" of each cluster (see main text), only approximately compensated by a self-coupling term $\tilde{\cal J}_{ii} \hat{b}_i^{\dagger} \hat{b}_i$, which is artificially decoupled \emph{\`a la} mean-field despite its local nature.     
 
 The rather annoying feature of the chemical potential/coupling matrix shift, imposed by the Hubbard-Stratonovich transformation, is strongly mitigated upon increasing the cluster size. Indeed the matrix ${\cal J}_{ij}$ of intercluster couplings acquires an increasingly sparse form, which reduces the absolute value of its most negative eigenvalues (so that the minimum $K$ to ensure positive definiteness of $\tilde{\cal J}$ is also reduced) -- yet $\Lambda_{\rm min}$ is found to saturate to a finite value in the infinite-cluster limit $\lambda\to 0$. On the other hand, the boundary nature of the chemical potential shift mitigates its effect for increasingly large clusters. Further discussion on the choice of $K$ is to be found in Sec.~\ref{s.eff}.   
 
\section{Cluster mean-field theory as saddle-point approximation to the bAFMC action}

The connection between cluster mean-field theory and the action governing the bAFMC approach goes beyond the zero-temperature limit discussed above. The minimum of the effective action ${\cal S}_{\rm eff}[\Psi]$ with respect to the auxiliary field at any temperature, defining the saddle-point approximation,  gives the equation
\begin{equation}
\frac{\delta}{\delta \Psi_i^*} S_{\rm eff} =  \beta \left [\sum_j X_{ij} \Psi_j - \frac{\delta F_c}{\delta \Psi_i^*} \right ] = 0
\end{equation}  
 where $F_c = -k_B T \log {\cal Z}_c$ is the single-cluster free energy. Clearly 
 \begin{equation}
 \frac{\delta F_c}{\delta \Psi_i^*} = \langle \hat b_i \rangle_c = \frac{1}{{\cal Z}_c} {\rm Tr} \left [ \hat{b}_i ~e^{-\beta\hat{\cal H}_c} \right]
 \end{equation}
 so that the saddle-point approximation to the bAFMC action produces the (modified) cluster mean-field condition 
 \begin{equation}
 \Psi_i = \sum_j \tilde{\cal J}_{ij} \langle \hat{b}_j \rangle_c
 \end{equation}
 at any finite temperature. This result shows that the bAFMC approach to bosonic quantum field theories surpasses the cluster mean-field approach by the inclusion of inter-cluster correlations -- but \emph{without} inter-cluster entanglement. In the $T=0$ limit the classical inter-cluster correlations described by the bAFMC approach disappear, so that bAFMC and (modified) cMF coincide, as found in the previous section. 
 
 \section{Effective couplings for the auxiliary field: spatial structure, chemical potential shift}
 \label{s.eff}
 
 The effective classical action for the time-independent auxiliary fields $S_{\rm eff}[\Psi]$ contains local intra-cluster terms ($-\log {\cal Z}_c$) as well as a non-local term with both intra- and inter-cluster couplings, with coupling matrix $X$. To understand the spatial structure of the couplings contained in the latter matrix, it is useful to start from the limit of clusters made of single sites only, in which case 
 \begin{equation}
  X_{ij} = \frac{1}{L^d} \sum_{\bm k} \frac{e^{i {\bm k}\cdot ({\bm r}_i - {\bm r}_j)}}{-e_{\bm k} + K}
  \label{e.coupling}
 \end{equation}
 where $e_{\bm k}$ is the dispersion relation given by the eigenvalues of the total hopping matrix $-J_{ij}$, namely the dispersion relation for the non-interacting limit of the model.    
 In the case of a hypercubic lattice with nearest-neighbor hopping $J$, $e_{\bm k} = -2J \sum_{a=1}^{d} \cos(k_a)$ (where the lattice spacing is taken as unity), and $K = 2dJ(1+\epsilon)$. As $-e_{\bm k}$ has a minimum for ${\bm k} = {\bm Q} = (\pi, \pi, ...)$ (leading to a maximum amplitude of the integrand), this wavevector dominates the integral in Eq.~\eqref{e.coupling}; shifting the integration variable to ${\bm q} = {\bm k}  - {\bm Q}$, and expanding around ${\bm q} = 0$, we obtain for $X_{ij}$ the form
 \begin{equation}
 X_{ij} \approx e^{i{\bm Q}\cdot({\bm r}_i-{\bm r}_j)}~\int \frac{d^d q}{(2\pi)^d} \frac{e^{i {\bm q}\cdot ({\bm r}_i - {\bm r}_j)}}{dJ (\xi^{-2} + q^2)}
 \end{equation}
 where $\xi \sim \sqrt{1/(2\epsilon)}$. The above integral is the Fourier transform of a Lorentzian, which decays as $\exp(-|{\bm r}_i - {\bm r}_j|/\xi)$ at large distance. Fig.~\ref{f.JSL} shows the characteristic decay of the coupling on the square and triangular lattice respectively. As expected, we observe that the exponential decay rate of the effective couplings is controlled by $\epsilon$; moreover at a fixed $\epsilon$ the decay length is shorter the bigger the clusters. The $X_{ii}$ term is always dominant and positive, bounding the amplitude fluctuations of the auxiliary field with a Gaussian distribution $\exp(-\beta X_{ii} |\Psi_i|^2)$. The off-site couplings $X_{ij}$ alternate in sign in the square lattice (because of the ${\bm Q}$ term), but are dominated by the negative nearest-neighbor term, favoring alignement of the phases of the auxiliary fields. In the frustrated triangular lattice, on the other hand, the couplings are all positive (as the maximum of the $e_{\bm k}$ dispersion relation is realized at ${\bm Q}=0$). The latter favors anti-alignement of the phases between neighboring auxiliary fields, which is of course frustrated by the lattice geometry.  
 
 \begin{figure}
 \includegraphics[width = \linewidth]{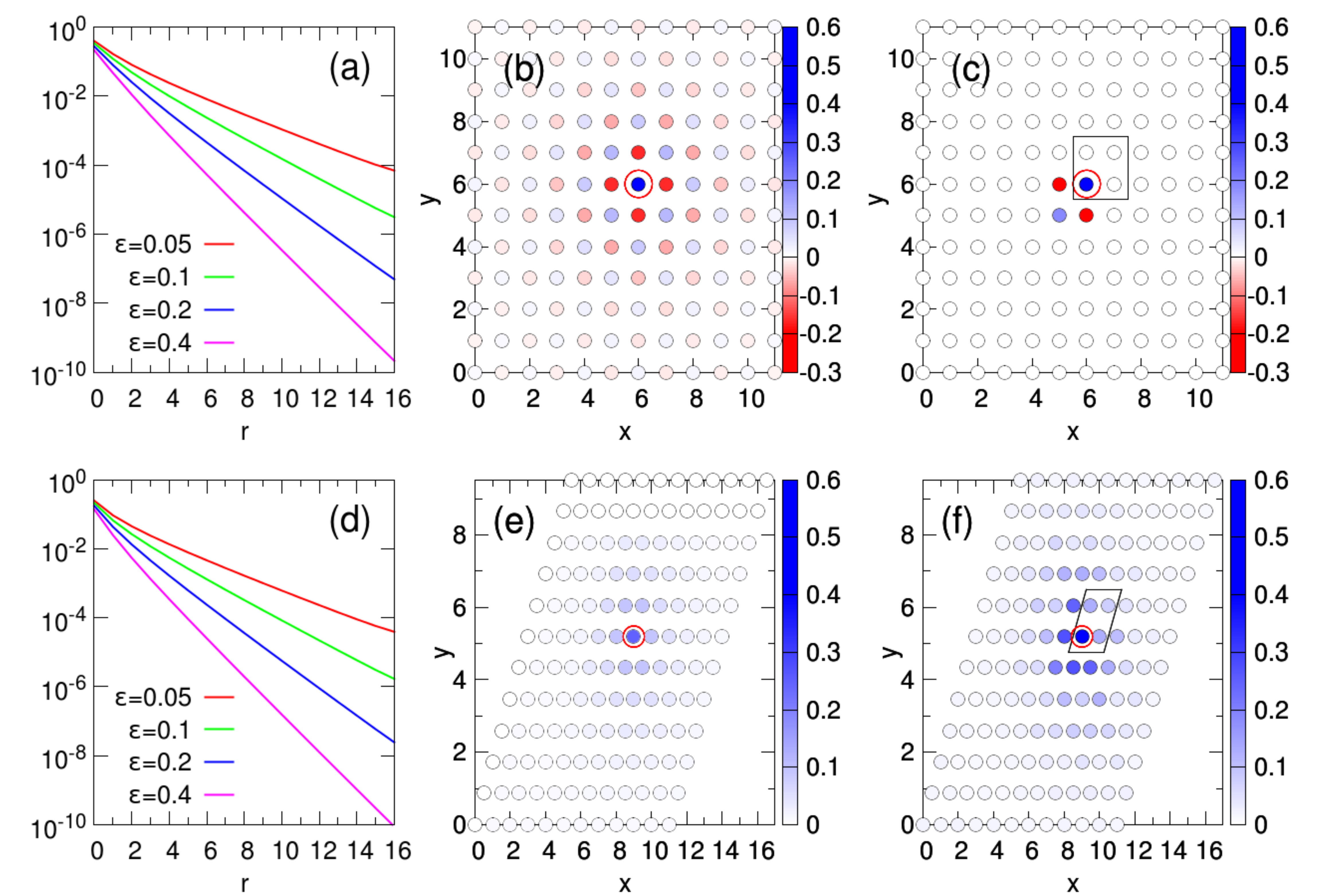}
	\caption{Effective auxiliary-field couplings $X_{ij}$. Upper row -- square-lattice couplings: (a) absolute value of the $X_{ij}$ couplings for ${\bm r}_j = {\bm r}_i+(r,0)$, and various values of $\epsilon$; (b) full structure of the coupling matrix for single-size clusters ($\epsilon = 0.05$); (c) same as in (b) for $2\times 2$ clusters. Lower row -- triangular-lattice couplings: (d) $X_{ij}$ couplings for ${\bm r}_j = {\bm r}_i+(r,0)$, and various values of $\epsilon$; (e) full structure of the coupling matrix for single-size clusters ($\epsilon = 0.05$) -- the encircled site is the $i$ site; (f) same as in (e) for one of the two inequivalent sites of rhombic clusters.}
	\label{f.JSL}
      \end{figure}

 From the previous example we see that the choice of the $K$ shift, parametrized by $\epsilon$, governs fundamentally the spatial structure of the couplings. At first sight, it appears that a natural choice for $\epsilon$ would be $\epsilon \gg 1$ in order to reduce the range of the effective couplings $X_{ij}$. At the same time, as discussed in the previous section the shift of the coupling matrix is only approximately compensated by that of the chemical potential in the cluster effective Hamiltonian, and therefore this would suggest to keep $\epsilon$ small in order to reduce this effect. Our choice for the simulations presented in this work is $\epsilon = 0.05$; we observe that, despite the approximation associated with the shift of the coupling matrix, the $\epsilon$ dependence of the simulation results  is rather moderate for $\epsilon$ in the range $0\div 1$ . Even if a specific choice of $\epsilon$ appears to be arbitrary for any fixed size of the the clusters, the infinite-cluster extrapolation ($\lambda\to 0$) must converge to the same limit regardless of the choice of $\epsilon$. In this sense, the value of $\epsilon$ affects the speed at which the extrapolation converge -- expected to be faster the lower $\epsilon$ --  as well as the convergence of each individual bAFMC simulation, as shorter-ranged couplings (obtained with a larger $\epsilon$) are generally expected to lead to a faster Monte-Carlo dynamics than long-ranged ones.  
 
  When working at a desired target filling $\langle n_i \rangle = \bar{n}$, one can use a further trick in order to reduce the dependence of the results on the cluster size, and hence accelerate the convergence towards the infinite-cluster limit. The simple trick is to \emph{adjust} the chemical potential $\mu$ appearing in the cluster effective Hamiltonian $\hat{\cal H}_c$ so that $\langle n_i \rangle(\mu) = \bar{n}$ for \emph{any} cluster size. This amounts in practice to redefine the chemical potential shift imposed by the Hubbard-Stratonovich transformation in such a way as to obtain the desired average density for \emph{every} cluster size. Without this trick the desired filling would instead only appear in the infinite-cluster limit. This is particularly convenient in the case of hardcore bosons, relevant for all the results presented in this paper. There the target average filling is $\bar n = 1/2$, which is easily achieved by taking a zero chemical potential in the effective cluster Hamiltonian regardless of the cluster size.

  \begin{figure}[ht!]
 \includegraphics[width = \linewidth]{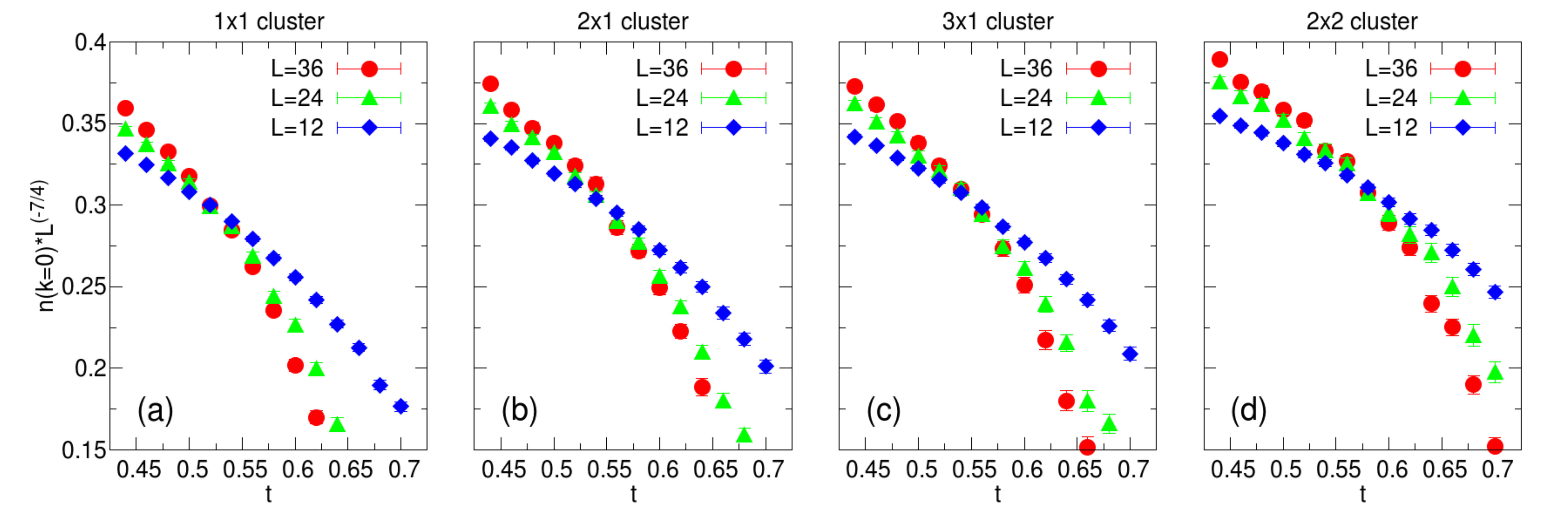}
	\caption{Scaling plots of the $n(k=0)$ peak in the momentum distribution for various cluster decompositions of hardcore bosons on the square lattice: (a) $1\times 1$ cluster; (b) $2\times 1$ cluster; (c) $3\times 1$ cluster; (d) $2 \times 2$ cluster.}
	\label{f.SLscaling}
      \end{figure} 
  
 \section{Bosonic auxiliary-field Monte Carlo: update algorithm and estimators}
 
 The bAFMC approach consists of a Monte Carlo simulation of the effective classical action $S_{\rm eff}[\Psi]$ for the complex lattice field $\Psi_i = |\Psi_i|^{i\theta_i}$. The phase-amplitude decomposition suggests that a minimal update scheme ensuring ergodicity involves local (single-site) phase and amplitude updates, which can be accepted or rejected with conventional Metropolis probabilities. Every such update requires to calculate the change in the partition function of the cluster containing the auxiliary field, as well as the variation of the action coming from the term containing the effective couplings  $X_{ij}$ (see Eq.~(6) of the main text). Given the exponentially decaying nature of the $X_{ij}$ couplings, the latter can be truncated to within some effective range $R$, so that the computational cost of an MC sweep attempting an update of each auxiliary field scales as $L^d \times [{\cal O}(d_H^{3n_c}) + {\cal O}(R^d)]$, where the first term comes from the cluster partition function ($d_H$ being the dimension of the local Hilbert space and $n_c$ the number of sites in the cluster), while the second one comes from the $X_{ij}$ coupling term.
 The simple single-site updates guarantee a good convergence of the results for the square lattice, while the same MC dynamics is more exposed metastable states in the case of the frustrated triangular lattice. For the latter it was necessary to carefully equilibrate the system via simulated annealing.  
 
  The statistical averages of local cluster operators $\hat O_c$ (namely containing field operators $\hat{b}_i, \hat{b}^{\dagger}_i$ with $i \in c$) are obtained as MC averages of estimators $O_c[\Psi]$ of the form
  \begin{equation}
  \langle \hat{O}_c \rangle = \langle O_c[\Psi] \rangle_{\rm MC} = \frac{\int {\cal D}[\Psi] ~O_c[\Psi] ~e^{-S_{\rm eff}[\Psi]}}{\int {\cal D}[\Psi] e^{-S_{\rm eff}[\Psi]}}
  \end{equation}
 where
 \begin{equation}
 O_c[\Psi] = \frac{1}{{\cal Z}_c}~{\rm Tr} \left[\hat{O}_c ~e^{-\beta \hat{\cal H}_c(\{\Psi_{i\in c},\Psi^*_{i \in c}\})}\right ]~.
 \end{equation}
 Observables involving operators associated to different clusters admit factorizable estimators, namely if $\hat{O} = \hat{O}_{c_1} \hat{O}_{c_2} ... \hat{O}_{c_m}$, then 
 $O[\Psi] = O_{c_1} O_{c_2} ... O_{c_m}$.

  \begin{figure}[ht!]
 \includegraphics[width = 0.8\linewidth]{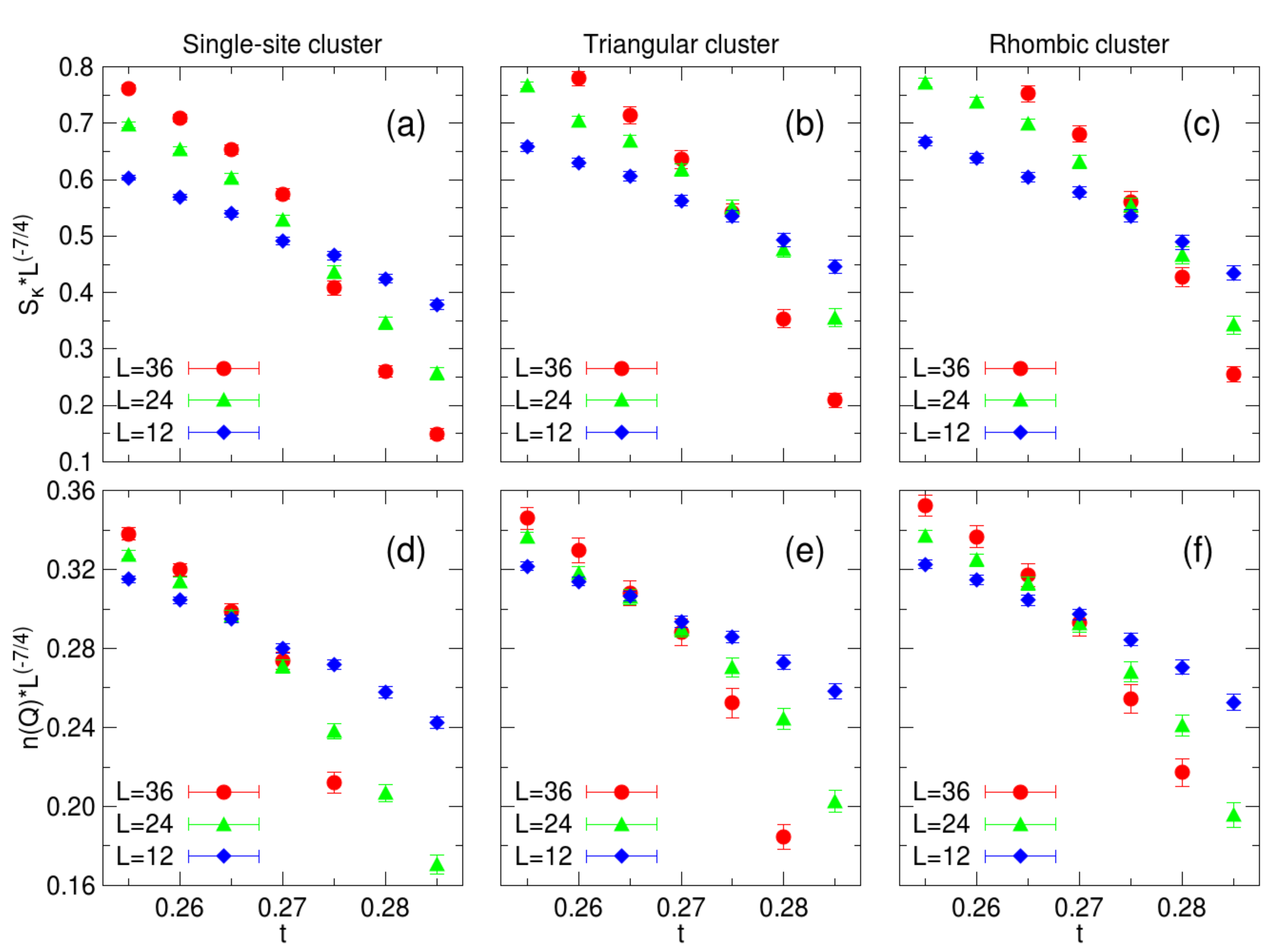}
	\caption{Scaling plots at the BKT and chiral transition of hardcore bosons on the frustrated triangular lattice. Upper row: $n({\bm Q})$ peak in the momentum distribution for various cluster decompositions: (a) single-site cluster; (b) triangular cluster; (c) rhombic cluster. Lower row: $S_{\kappa}$ peak in the chiral structure factor for various cluster decompositions: (d) single-site cluster; (e) triangular cluster; (f) rhombic cluster.}
	\label{f.TLscaling}
      \end{figure}
 
 \section{Finite-size scaling analysis of hardcore bosons on the square and triangular lattice}
In this section, we provide all the data used for the finite-size scaling analysis of the transitions of hardcore bosons on the square and triangular lattice. Fig.~\ref{f.SLscaling}, referring to the square lattice, shows that a convincing estimate of the BKT transition can be obtained via the scaling $n(k=0)\sim L^{7/4}$ for sizes $L=$12, 24 and 36 for all cluster decompositions, and that the estimated critical temperature increases gradually as the cluster size is increased.  Fig.~\ref{f.TLscaling} shows a similar observation for the BKT transition on the triangular lattice. Moreover the chiral transition is also analyzed by looking at the chiral structrure factor $S_{\kappa} = L^{-2} \sum_{\triangle,\triangle'} \langle  \kappa_{\triangle} \kappa_{\triangle'} \rangle$. For single-site and triangular clusters, the sum has been restricted to $L^2/3$ up-pointing triangles $\triangle, \triangle'$ regularly tiling the triangular lattice (see Fig.~\ref{f.TLdecomposition}(a)); in the case of rhombic clusters it is restricted to $L^2/12$ up-pointing triangles fully contained in the $L^2/4$ rhombi (see Fig.~\ref{f.TLdecomposition}(b)), and multiplied by a factor of 4 to compare with the other cluster decompositions. We observe that the Ising critical scaling $S_{\kappa} \sim L^{7/4}$ allows to consistently estimate the chiral critical temperature for every cluster decomposition, and that such an estimate lies systematically above the one for the BKT transition.   
 \newpage
 
 \begin{figure}[ht!]
 \includegraphics[width = 0.7\linewidth]{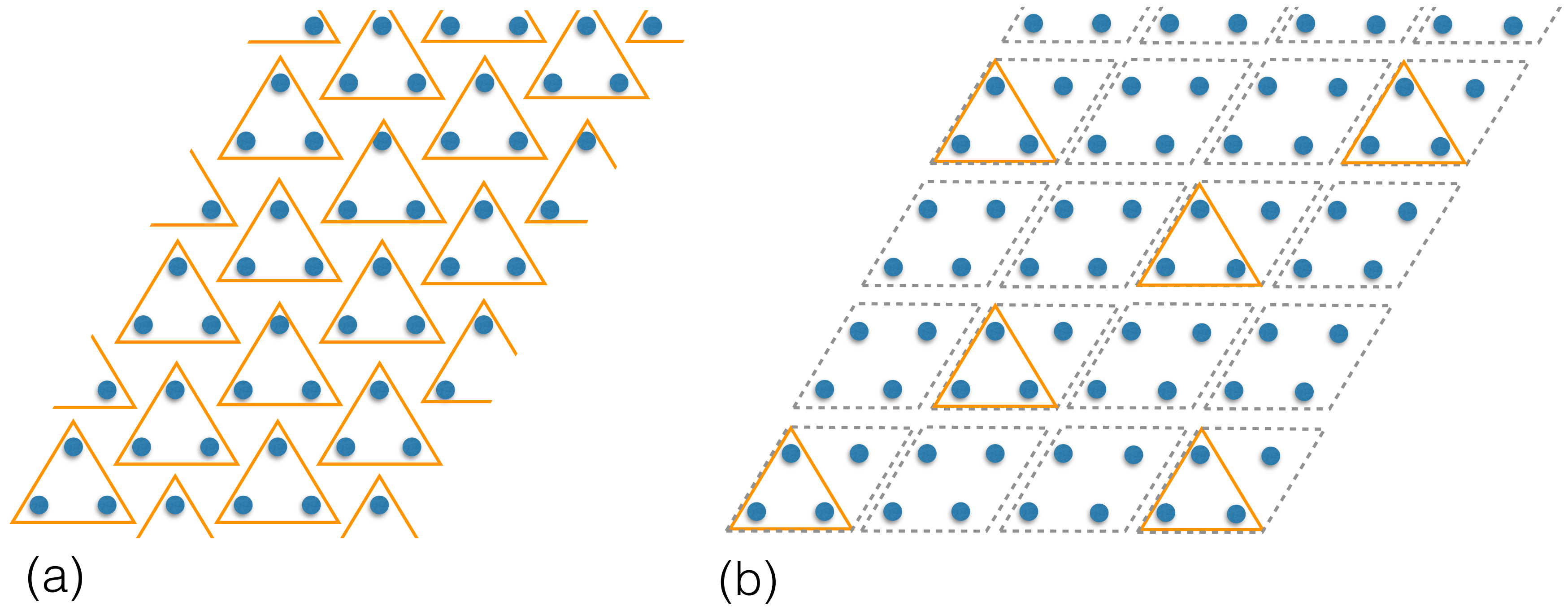}
	\caption{Cluster decompositions for the triangular lattice used in this work: (a) triangular clusters; (b) rhombic clusters. In both panels, the orange triangles indicate the plaquettes used to calculate the chiral structure factor $S_{\kappa}$.}
	\label{f.TLdecomposition}
      \end{figure}  
 
\end{widetext}

\bibliography{AFMC}

\end{document}